\documentclass{aa}  
\usepackage{natbib}
\bibpunct{(}{)}{;}{a}{}{,}
\AtBeginDocument{\nolinenumbers\let\linenumbers\relax}
\usepackage{graphicx}
\usepackage{txfonts}
\begin{document} 

\nolinenumbers

\title{Effect of Turbulence and Inhomogeneities on the Propagation of Kinetic Scale Waves}

   \author{T.A. Bowen
          \inst{1}
         }

   \institute{Space Sciences Laboratory, University of California,7 Gauss Way, Berkeley, CA 94720\\
   \email{tbowen@berkeley.edu}}

  \abstract
   {Electromagnetic ion-scale waves are well known to occur in the solar wind. Eikonal (WKB) like effects may play an important role in the interplay between the propagation and local kinetic plasma dynamics of ion cyclotron waves.}
   {This Letter aims to explore the interplay of ICW propagation with turbulence and local kinetic plasmas using ray tracing equations for ICWs in a moving medium, in which turbulent gradients are present.}
   {Parker Solar Probe observations of ICWs in the solar wind are studied using ray equations in a moving medium. Gradients in the dispersion relation responsible for WKB evolution of ICW are constrained and compared with a Parker spiral model.}
   {The empirical gradients, which contain turbulent motions, greatly enhance the efficiency of WKB effects. However, the typical refraction rates are significantly slower than expected quasilinear evolution of the observed wave populations, which suggests that local kinetic physics dominates over nonlinear expansion effects.} 
{Turbulence can effectively increase WKB effects due to gradients caused by nonlinear interactions. In the solar wind these gradients are not sufficient to substantially change an individual ion-scale ICW over its lifetime. Incorporating kinetic effects of expansion may help advance the understanding of generation and absorption of ICW and their role in heating in the solar wind.}

   \maketitle
%

\section{Introduction} Recent observations from Parker Solar Probe have brought renewed interest in the fundamental physics of ion cyclotron waves and their role in dissipating turbulent energy. Observations of the solar wind often contain ion-cyclotron resonant waves \citep{Jian2010}, with the fractional occurrence of these waves increasing with proximity to the Sun \citep{Boardsen2015,Liu2023}. The existence of these waves has been linked to imbalanced intervals, in which the turbulence resembles Alfvén waves \citep{Bowen2024a,Zhao2026}. Cyclotron resonant processes have long been a candidate mechanism for collisionless dissipation \citep{Marsch1982a,Isenberg2007,Isenberg2011}; however, the tendency of turbulence to form anisotropic perpendicular structure that is elongated along the mean field poses difficulty for the fluctuations to reach parallel scales sufficiently small for ion cyclotron resonance to occur \citep{Horbury2012,Chen2012}. Despite this known difficulty in direct formation of parallel propagating ICW from turbulence, ion-scale waves in the solar wind show circular polarization indicating parallel propagation \citep{Jian2010,Wicks2016,Bowen2020a}, and heating rates are similar to turbulent dissipation rates \citep{Bowen2022,Bowen2024b,Shankarappa2024}. The growth and damping of these waves can occur without turbulence via kinetic instabilities \citep{Gary1993,Klein2021}. The regular observation of kinetic distribution functions that indicate cyclotron resonant processing \citep{Marsch2001,He2015} suggests an important role in heating in the solar wind. Furthermore, there is significant evidence that ICW shape turbulence at sub-ion kinetic scales \citep{Bowen2024a}, highlighting physical connections between the turbulence and ion-scale wave populations. 

The physics of imbalanced turbulence remains relatively poorly understood, though it is increasingly recognized to be important in shaping a range of astrophysical environments \citep{Lithwick2007,Wong2024}. Imbalanced turbulence is thought to be subject to a barrier at kinetic scales \citep{Meyrand2021} that results from the tendency for generalized helicity to cascade in opposite directions--a forward cascade at magnetohydrodynamic (MHD) scales and an inverse cascade at sub-ion scales--near the ion-gyroscale. Particle in cell simulations, e.g. \cite{Squire2022}, captured the helicity barrier, resulting in increased amplitude and decreased anisotropy of the fluctuations in the MHD-range inertial turbulence. These fluctuations can then heat the plasma via either stochastic heating \citep{Chandran2010,Bowen2025c} or oblique ion cyclotron resonance \citep{Chandran2010b} depending on plasma $\beta$ and the level of imbalance. In the simulations of \cite{Squire2022}, oblique ICW (OICW) heating eventually drives parallel ICW (PICW) unstable. 

\begin{figure}
    \centering
    \includegraphics[width=\columnwidth]{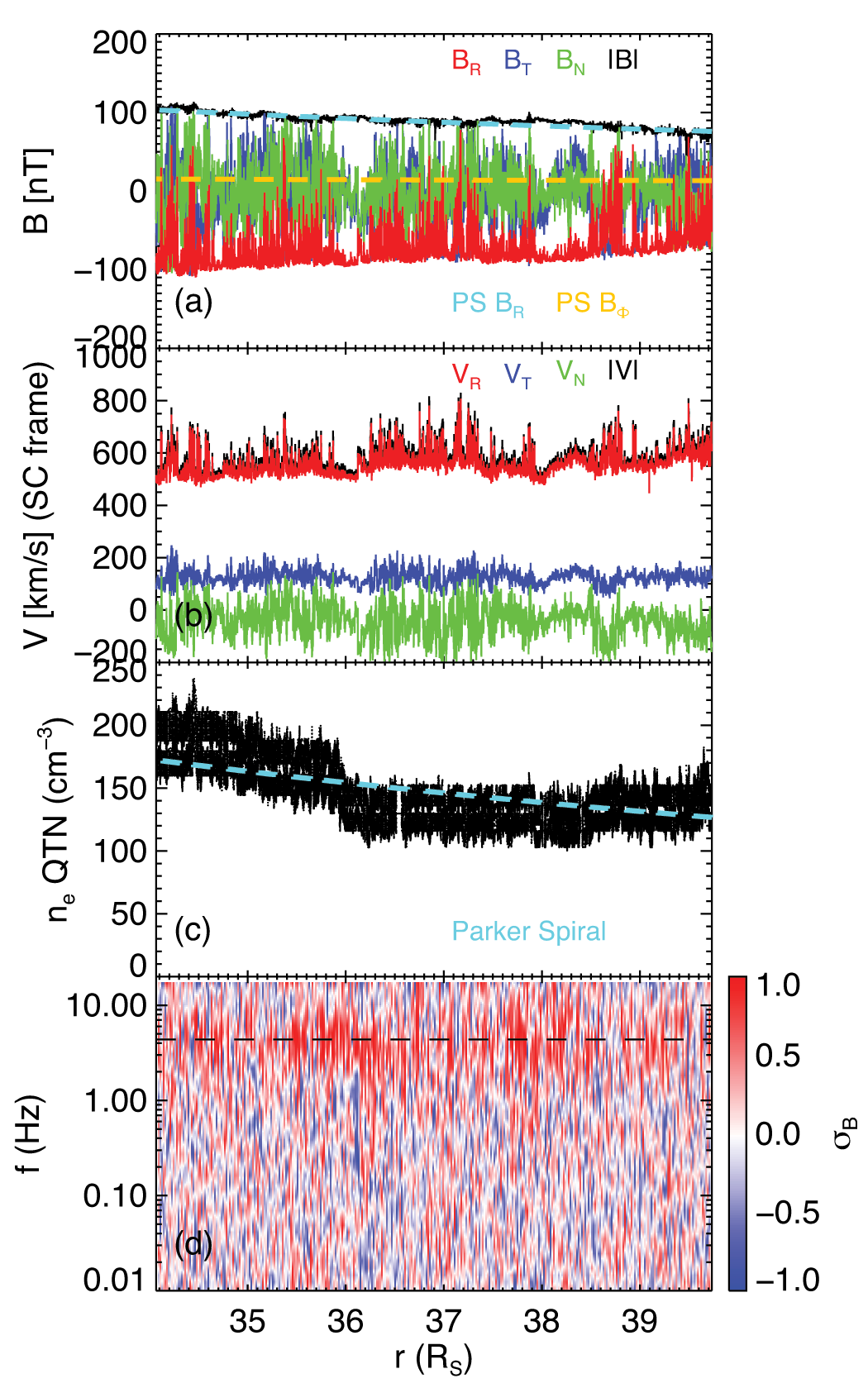}
    \caption{a) PSP magnetic field data in RTN coordinates with a Parker spiral model superimposed. b) PSP velocity data in the spacecraft (SC) frame. c) PSP electron density from quasithermal noise (QTN), upper and lower bounds including 8\% fractional uncertainty are shown \citep{Pulupa2017,Moncuquet2020}. d) Reduced magnetic helicity $\sigma_B$ from wavelet transform, red is left hand polarized with $\sigma_B=1.$}
    
\end{figure}
 \begin{figure}
     \centering
     \includegraphics[width=\columnwidth]{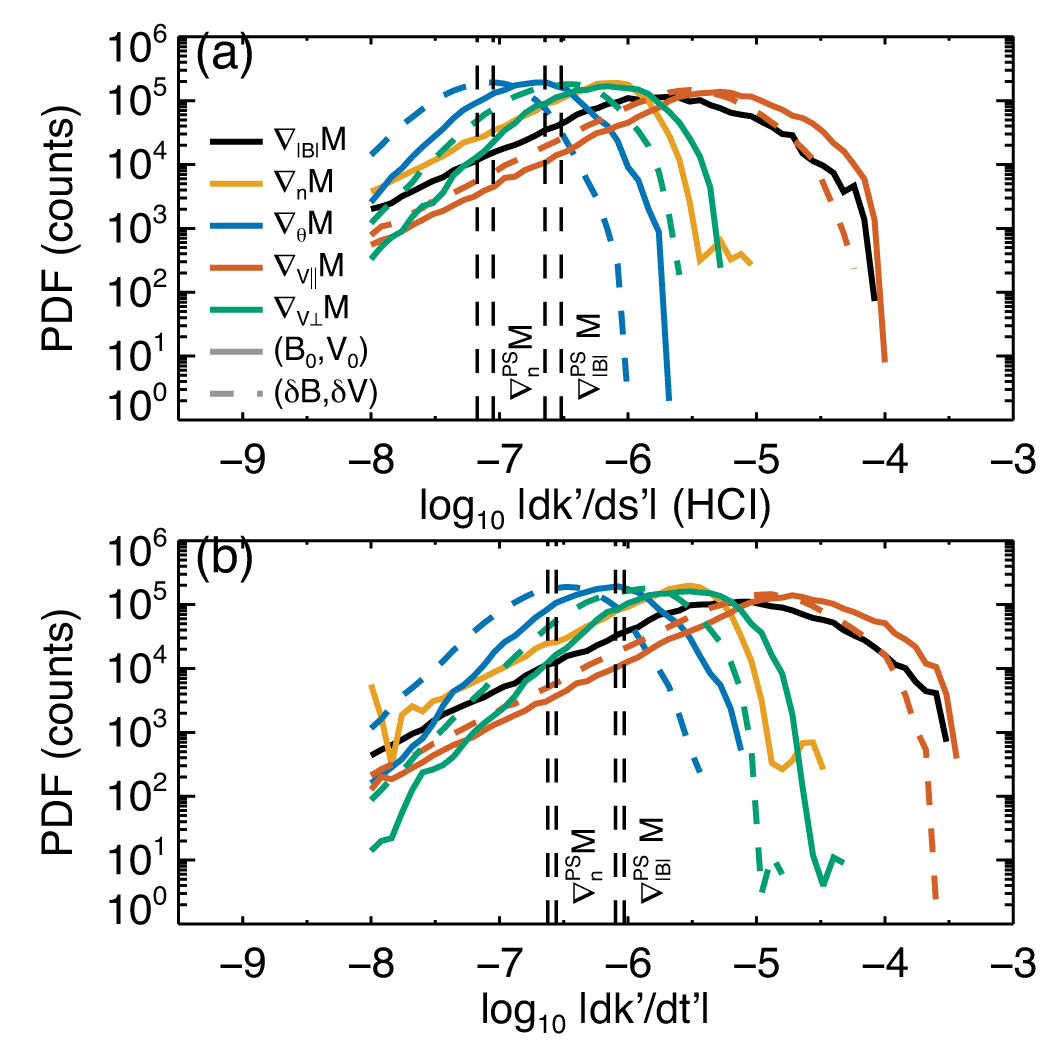}
     \caption{a) Normalized $dk'/ds'$ for measurements from PSP, which include turbulent gradients. Each term affecting WKB propagation is shown in a separate color (see legend). Vertical dashed lines show the dominant Parker spiral terms $\nabla\mathcal{M}_n$ and $\nabla\mathcal{M}_{|B|}$. b) Same as the top panel but for $dk'/dt'$, the change in normalized wavenumber per cyclotron time. Vertical lines show the Parker spiral terms.}
     
 \end{figure}

While the helicity barrier provides a novel and self consistent framework for explaining heating in imbalanced turbulence  that is roughly consistent with observations \citep{Bowen2024a,McIntyre2025,Panchal2025,Zhao2026}, there is some difficulty in reconciling the generation of PICW, which are driven unstable by the OICW resonance \citep{Chandran2010b}, with observed damping rates \citep{Bowen2022, Bowen2024b}. Recently, \cite{Yerger2026} proposed a mechanism, termed cyclotron breaking, in which a process of continuous emission and absorption of PICW occurs due to gradients in the background field; specifically,  PICW that are unstable at a certain solar radius propagate into a region of space, refract off background gradients and then become stable and damp, thus resulting in net heating. This cyclotron breaking process has supporting evidence in the bands of emission and absorption present in observations \citep{Bowen2024b}.
\section{Ray tracing in a turbulent medium}

In this Letter, ray tracing formalism is applied in a moving medium, including gradients introduced by turbulence, to study the evolution of ICW in the solar wind within the eikonal approximation \citep{Hazeltine2004}.
The plasma frame dispersion relation is $\mathcal{M}(\omega,\mathbf{k},\mathbf{r})=0$ with $\omega=\Omega(\mathbf{k},\mathbf{r})$. Defining normalized units $\omega'=\omega/\Omega_p$ and $k'=kd_p$ and $\theta$ as the angle between $\mathbf{B}$ and $\mathbf{k}$, for cold-plasma ion cyclotron waves
\begin{align}
\mathcal{M}(\omega,\mathbf{k},\mathbf{r})= \frac{{k'^4 \cos^2\theta (1-\omega'^2)} -{\omega'^2k'^2(1+\cos^2\theta)} +{\omega'^4}}{{\omega'^4 (1-\omega'^2)}}=0,
\end{align}
 \citep{stix1992waves}. The gradient in $\mathcal{M}$ is given as
\begin{align}
\nabla{\mathcal{M}}=-\frac{\partial{\mathcal {M}}}{\partial{\omega'}}\frac{\omega'}{|B_0|}\nabla |B_0| - \frac{\partial{\mathcal {M}}}{\partial{k'}}\frac{k'}{2n_p} \nabla n_p +\frac{\partial\mathcal{M}}{\partial{\theta}}\nabla\theta. \label{eq:gradm}\end{align}
each of the terms is outlined in the Appendix.
In the solar wind, gradients in background heliospheric structure consist of large scale variations governed by the Parker spiral (PS) 
\begin{align}\mathbf{B}_{ps}(R)=\frac{B_{R_0}{R_0^2}}{R^{2}}\hat{R} -{B_{R_0}\frac{\Omega_\odot}{v_{sw0}} \frac{R_0^2}{R}}\text{sin}\Theta\hat{\Phi} \label{eq:parker_spiral}.\end{align}
For simplicity the PS density $n_e$ will be considered to follow an $r^{-2}$ spherically decaying profile and the flow $v_{sw0}$ will be taken as constant in the $\hat{R}$ direction. WKB evolution is typically considered in the context of large scale heliospheric structure \citep{Heinemann1980,Hollweg1990}. However, for small, kinetic scales, turbulent fluctuations provide a local field and flow gradients that deviate from background PS profile, which as shown  here, are important to the dynamics of ion-scale waves and their kinetic interactions with the plasma.

 Ray equations in the heliocentric inertial (HCI) frame, with flow moving at $\mathbf{v}$ and a Doppler shifted $\omega_{HCI}=\omega +\mathbf{k}\cdot\mathbf{v}$ can be derived as
 \begin{align}
  \frac{d \mathbf{r}_{HCI}}{dt}=\frac{\partial{\Omega}}{\partial \mathbf{k}}+\mathbf{v}=-\frac{\partial{\mathcal M_{HCI}}/\partial{\bf k} }{\partial{\mathcal M_{HCI}}/\partial\omega}\label{eq:drdt_hci}\\
 \frac{d \mathbf{k}_{HCI}}{dt}=-\nabla\Omega-(\nabla \mathbf{v})\cdot\mathbf{k}=\frac{\partial{\mathcal M_{HCI}}/\partial{\bf r} }{\partial{\mathcal M_{HCI}}/\partial\omega} \label{eq:dkdt_hci}
 \end{align}
 where $\mathcal{M}_{HCI}(\omega,\mathbf{k},\mathbf{v},\mathbf{r},t)=0$ is the HCI frame dispersion and \begin{align} \frac{d\mathbf{k}_{HCI}}{ds}=\frac{-\nabla\Omega-(\nabla \mathbf{v})\cdot\mathbf{k}}{|\frac{\partial{\Omega}}{\partial \mathbf{k}}+\mathbf{v}|}=-\frac{\nabla \mathcal{M}_{HCI}}{|\partial \mathcal{M}_{HCI}/\partial{\mathbf k}|}. \label{eq:dkds_hci}\end{align}
 Equation \ref{eq:dkds_hci} represents the change in wave-vector with distance, when including a turbulent background flow. Similarly, the change in time is given by Eq. \ref{eq:dkdt_hci}. Subsequently, the notation $\nabla_x \mathcal{M}$ refers to the contribution in $\nabla \mathcal{M}$ resulting from gradients in $x$.

Figure 1(a) shows Parker Solar Probe observations on November 18th, 2021, over which the spacecraft descended from 
39.7 to  34.1 $R_\odot$ and left-handed waves were previously documented \citep{Bowen2024b}. Measurements from PSP FIELDS provides magnetic field \citep{Bale2016} and density \citep{Moncuquet2020}, while the solar wind velocity is measured from SWEAP-SPAN-I \citep{Kasper2016,Livi2022}. A Parker spiral model is fit to the data.

Circular polarization is determined
 by the {reduced} magnetic-helicity $\sigma_B(f,t)$ \citep{HowesQuataert2010}.
Figure 1(d) shows reduced magnetic helicity determined from a wavelet transform \citep{Bowen2020a,Bowen2024b}. Left handed intervals are clearly evident at frequencies corresponding to the ion-inertial length $2\pi f_{di}=|V_{sc}|/d_p$, with $f_{di}\approx 4$ Hz and $V_{sc}\approx600$ km/s, where $V_{sc}$ is the solar wind speed measured in the spacecraft frame. The wavelet transform spans from 17.5 Hz to 0.005 Hz, corresponding to $\sim$ 0.06 to $\sim 200$ seconds. The median solar wind speed in the HCI frame is 567 km/s and with a median proton inertial length of $20$ km, results in a range of 0.25 to 800 $d_p$, where $d_p=c/\omega_p$ and $\omega_p$ is the proton plasma frequency. Gaussian smoothing is performed using the modulus of each Morlet wavelet to give a scale dependent $\mathbf{B}_0(f,t)$ corresponding to the vector coefficients at each wavelet scale $\mathbf{B}_1(f,t)$.

 WKB-like refraction occurs for gradient scales much longer than the wave-packet. For a Morlet 6th order wavelet, the full width half max of the wavelet contains approximately two full wave periods, for $\lambda/L\sim 0.1$ or $kL\sim20\pi$, with $L$ a characteristic gradient scale, gradients on scales 4-5 times larger than the 6th order Morlet wavelet should meet ``slowly varying'' conditions. Comparison at average scales over the lifetime of a wave is likely more meaningful, as these are the net gradients seen by the wave. Solar wind plasma is often in marginal stability with ICWs \citep{Marsch1982a,Bowen2022}, such that $|\gamma/\omega| \approx$ $10^{-1}$ to $10^{-3}$. Accordingly, gradients at 100-1000 $d_p$ are most important in interplay between WKB and kinetic processes. The subsequent analysis studies the interplay of waves at the ion inertial scale $k{d_p}\approx 1$, corresponding to a frequency of 4 Hz, with gradients at $\approx 100d_p$, at $\approx 0.034$ Hz. Both secular, nonstationary, fluctuations at the gradient scale are considered (e.g. $B_0$), as well as mean-zero turbulent fluctuations (corresponding to $B_1$), that maintain gradients that vary over scales larger than $kd_p$. 
 \begin{figure}
    \centering
    \includegraphics[width=\columnwidth]{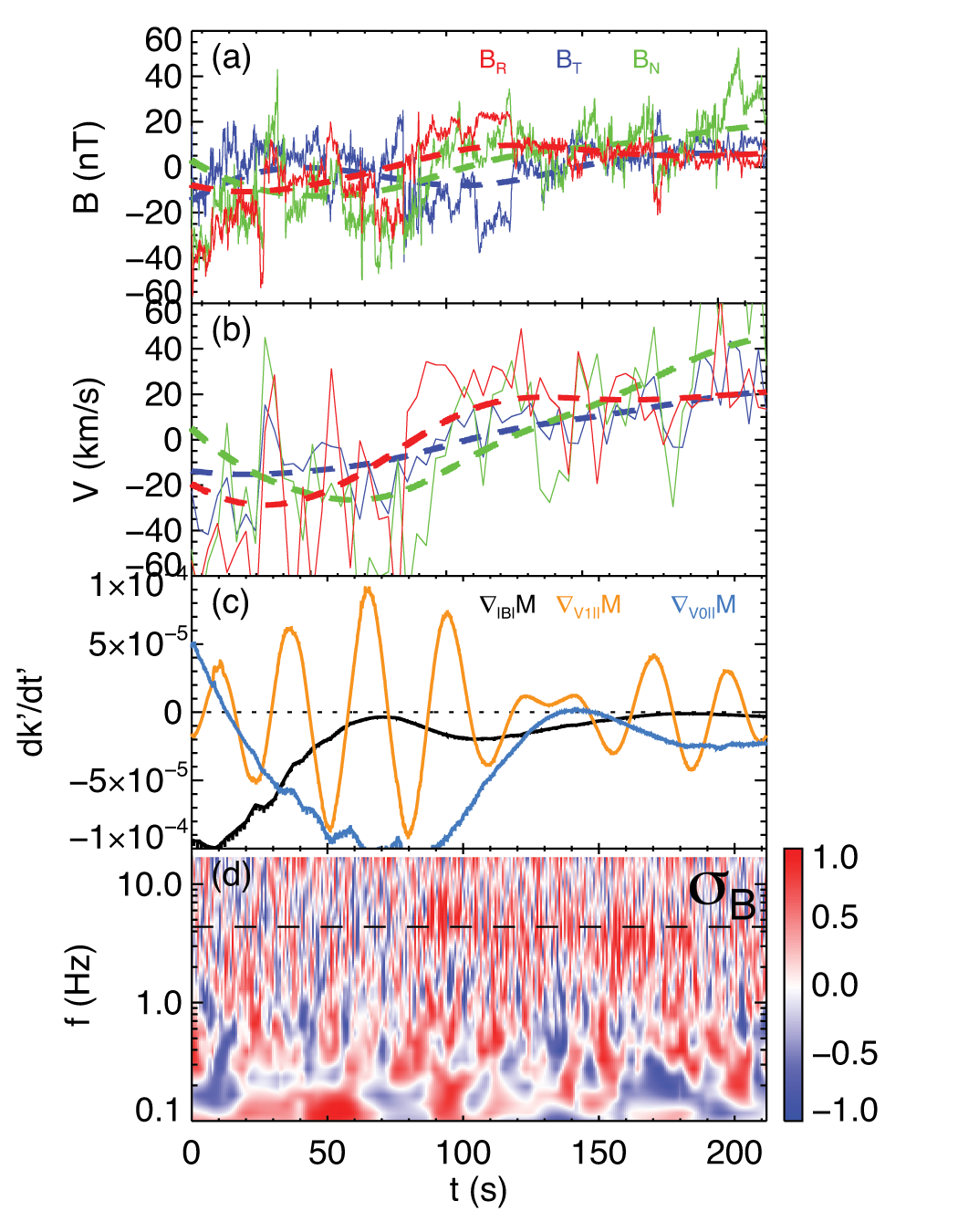}
    \caption{Interval starting 2021-11-18/08:01:46, with (a) mean subtracted magnetic field, (b) mean subtracted velocity (c) gradients in $\nabla_{|B|} \mathcal{M}$ (black) $\nabla_{v_{\parallel 0}}\mathcal{M}$ (blue) $\nabla_{v_{\parallel 1}}\mathcal{M}$ (orange); d) shows $\sigma_B$ at this time. Smoothed quantities (dashed) are at the 100$d_p\approx0.034$ Hz scale.}
    
\end{figure}

Single point observation effects complicate measurement of gradients in the solar wind, and evaluating the actual change in wave-vector cannot be performed without a fully defined vector field. However, measured gradients can provide a baseline understanding of the most important effects. The proxy of the full $\nabla| B_0|$ is determined by the time derivative of $|{B}_0(f,t)|$ sampled in the spacecraft frame, assuming the Taylor hypothesis, \begin{align}\nabla |B_0|\approxeq \frac{d}{dt}|\mathbf{B}_0(f,t)|/V_{sc} \label{eq:grad_magb}.\end{align} This term is the first right-hand-side term of Eq.~ \ref{eq:gradm}. The gradient of the density is similarly calculated as
\begin{align}
\nabla n_p\approxeq \frac{1}{{V}_{sc}}\frac{d}{dt}{n}_p(f,t)  \label{eq:grad_dens}\end{align}
The gradient in propagation angle, $\nabla \theta$ is estimated by identifying with $\cos\theta$= $\hat{b}\cdot\hat{k}$. Such that $\nabla \theta= \frac{-1}{\sin \theta}[(\nabla \hat{b} )\cdot\hat{k}+(\nabla\hat{k})\cdot\hat{b}].$ The spatial gradient in $\hat{k}$ at fixed wavenumber is zero and thus $(\nabla\hat{k})\cdot\hat{b}$ is dropped \citep{Hazeltine2004}. Then $\nabla \theta  =  (-\text{sin}\theta)^{-1}  (\nabla\hat{b})\cdot\hat{k}$.  Letting 
$\hat{k}= \text{cos}\theta\hat{b} + \text{sin}\theta \hat{u}$, with $\hat{u}$ an arbitrary perpendicular vector, $\nabla \theta= -(\nabla \hat{b})\cdot\hat{u}$. The term $(\nabla \hat{b})\cdot \hat{b}$ is by definition zero such that purely parallel waves are not affected. To include effects of $\nabla \theta$, a propagation angle of $2^\circ$ is used.

The $\nabla \theta$ term can be approximated by taking the derivative of each vector component of $\mathbf{B}(f,t)=\mathbf{B}_0(f,t)+\mathbf{B}_1(f,t)$; the vector derivatives $d \mathbf{B}_0/dt$ and $d \mathbf{B}_1/dt$ (again invoking the Taylor hypothesis) are then projected onto $\hat{B_0}$ to find the parallel component, which is then subtracted to find the contributions perpendicular to $\mathbf{B}_0$, while normalizing to the magnitude gives the change in field direction. Specifically, \begin{align} (\nabla\hat{b})\cdot\hat{u}\approx  \frac{1}{{V}_{sc}}\frac{d\mathbf{B}/dt-(d \mathbf{B}/dt\cdot\hat{b})\hat{b}}{|B|}\label{eq:curve},\end{align} where the contributions from $\mathbf{B}_1$ and $\mathbf{B}_0$ are separately considered. There is no a priori information on the direction of $\hat{u}$, such that $\hat{u}$ is along the gradient, maximizing the term.


The aforementioned terms provide a constraint on ray tracing for ICW (and a framework for other dispersive waves) in the plasma frame. However, in a moving medium with gradients in the flow speed, evaluation of the terms in Equation \ref{eq:dkds_hci} is necessary. Following the framework used in estimating $\nabla \theta$, the  $(\nabla \mathbf{v})\cdot\mathbf{k}$ are estimated as $\nabla \mathbf{v}\cdot(k\text{cos}\theta\hat{b} +k\text{sin}\theta \hat{u}) $ and $\mathbf{v}$ is broken into a slowly varying $\mathbf{v}_0$ and mean zero $\mathbf{v}_1$. Each of the gradients in $\mathbf{v}_0$  and $\mathbf{v}_1$ are estimated from the Taylor hypothesis and projected into field-parallel and perpendicular contributions:
\begin{align}\label{eq:flow_par}
     k\cos\theta\,(\nabla \mathbf{v})\cdot\hat{b}&\approx k\cos\theta \frac{\hat{b}}{{V}_{sc}}\frac{d\mathbf{v}}{dt}\cdot\hat{b}\\
     k\sin\theta\,(\nabla \mathbf{v})\cdot\hat{u}&\approx k\sin\theta \frac{1}{{V}_{sc}}\left(\frac{d\mathbf{v}}{dt}-\hat{b}\frac{d\mathbf{v}}{dt}\cdot\hat{b}\right)\label{eq:flow_perp}.
\end{align} 

Figure 2(a) shows approximate estimates of ${d{\mathbf{k}'_{HCI}}}/{ds'}$, in normalized units understood as change in wavenumber (in $kd_p$) per length traveled (in units of $d_p$) for each term in Equations \ref{eq:grad_magb}-\ref{eq:flow_perp}. Mean zero and secular gradients are considered for Equations, \ref{eq:curve}- \ref{eq:flow_perp}. The values for the Parker spiral model are similarly plotted. Figure 2(b) shows estimates for ${d{\mathbf{k'}_{HCI}}}/{dt'}$ in units of change in wavenumber (in $kd_p$ per $\Omega_p$). The results show that the WKB effects are greatly increased when incorporating turbulent flows in the solar wind. The dominant term in the WKB evolution corresponds to the change in magnitude in $|B|$, $\nabla_{|B|}\mathcal{M}$, (Eq.~\ref{eq:grad_magb}) as well as terms in Eq.~ \ref{eq:flow_par}, corresponding to gradients in the turbulent flows that are parallel to the mean field (and wave vector propagation) direction, terms ($\nabla_{v_{\parallel 0}}\mathcal{M}$, $\nabla_{v_{\parallel 1}}\mathcal{M}$ in Fig. 2 \& 3). Interestingly, these gradients both occur with $dk'/dt' \ll10^{-4}\Omega_p$, indicating that the effects of WKB expansion should be relatively weak even in cases of marginal stability where quasilinear processes are at all efficient (i.e. growth/damping rates of $\gamma\sim10^{-3}$). Figure 3 shows a small cut around 08:01:46 with the magnetic field and velocity plotted in panels a and b, and the dominant $\nabla \mathcal{M}$ terms in panel c.

\section{Discussion}

WKB like processes have been widely used to understand the evolution of waves in the expanding solar wind at MHD scales \citep{Heinemann1980,Hollweg1990}. Similarly, the dynamics of kinetic waves in an expanding plasma has been invoked in models of coronal heating \citep{Schwartz1981,Matthaeus1999,Cranmer2000,Leamon2000}. \cite{Yerger2026}  proposed that locally unstable cyclotron waves propagate into regions where they become stable, resulting in net heating.

PSP measurements constrain turbulent gradients present in the solar wind that affect the efficiency of WKB processes. The results indicate that inertial range Alfvénic turbulence provides important background gradients that increase the efficiency of WKB (in comparison to the background Parker spiral) via perturbation of local fields seen by PICW. However, WKB effects seem to only affect PICWs that exist for significantly longer than $10^4 \Omega_p^{-1}$. For these long lived waves it is important to simultaneously consider expansion effects  \citep{Chew1956} on particle distributions. Eikonal evolution depends on the cumulative change $\Delta k'$ over its lifetime. For mean-zero gradients (e.g. $\mathbf{v}_1$), $\Delta k'$ 
grows diffusively via a random walk. However, gradients in $|\mathbf{B}_0|$ and $\mathbf{v}_0$ remain coherent over
$\sim 100\,d_p$, resulting in secular drift $\Delta k'$. The
secular case is therefore an upper bound, as diffusion slows down the eikonal evolution.

Given that quasilinear growth and damping rates are significantly faster than refraction, it seems that local wave-particle interactions are primary to the nonlinear expansion effects on the waves. The relative difference in scales suggests that local instability analysis, often used to understand kinetic processes in the solar wind \citep{Klein2021,Mcmanus2024,Martinovic2026}, is applicable.
These results may complicate the theory of cyclotron breaking \citep{Yerger2026}, which relies only on Parker spiral gradients to drive WKB effects of PICW. The WKB effects enabled by turbulence can couple PICW back into the OICW via gradients in the background turbulence. It is possible that refraction generated obliquity helps maintain marginal stability. Specifically, as the equilibrium contours for OICW are innately steeper than PICW \citep{Chandran2010b,Yerger2026}, the generation of slight obliquity from WKB enables their reabsorption by a plasma in PICW-equilibrium, and the subsequent re-emission of PICW. This quasi-equilibrium may act to maintain the marginally stable PICW state regularly observed in the solar wind \citep{Marsch2001,He2015,Bowen2022}. Further investigation into cyclotron breaking in the presence of expanding plasma in a turbulent background is needed to fully understand this process.

\begin{acknowledgements} This work was supported by NASA grant 80NSSC24K0272. The author acknowledges the use of AI tools in validating derivations, proofing text, and generating code for figures under oversight of the author. No generative AI was used to compose the manuscript or to process or analyze data. All derivations were performed by the author's hand.\end{acknowledgements}
\bibliographystyle{aa} 
\bibliography{prl} 

\begin{appendix}
\section{Ray Tracing}
Following \cite{fitzpatrick2022plasma}, the ray tracing equations without a moving medium are
\begin{align}
\frac{{d}\mathbf{r}}{dt}=\frac{\partial{\Omega}}{\partial \mathbf{k}}=-\frac{\partial{\mathcal M}/\partial{\bf k} }{\partial{\mathcal M}/\partial\omega},\\
\frac{{d}\mathbf{k}}{dt}=-\nabla\Omega=\frac{\partial{\mathcal M}/\partial{\bf r} }{\partial{\mathcal M}/\partial\omega},\\
\frac{{d}\Omega}{dt}=\frac{{d}\omega}{dt}=-\frac{\partial{\mathcal M}/\partial{t} }{\partial{\mathcal M}/\partial\omega}
\end{align}
In the frame with mean zero flow, it is then possible to approximate the change in wavenumber with a scalar distance \begin{align}
\frac{d\mathbf{k}}{ds} = -\nabla \Omega/|{\partial{\Omega}}/{\partial \mathbf{k}}|=-\frac{\nabla\mathcal{M}}{d_p\partial{\mathcal M}/\partial k'}.
\end{align} The gradient in $\mathcal{M}$ is broken up term by term, where it is useful to define polynomials in $k'$ and $\omega'$ as \begin{equation} {Ak'^4+Bk'^2 +C=0}\end{equation} with 
$A=\cos^2\theta(1-\omega'^2)$, $B=-\omega'^2(1+\cos^2\theta)$ , and $C= \omega'^4$
as well as  \begin{equation} {a\omega'^4+b\omega'^2 +c=0}\end{equation} 
with $a=1$, $b=-k'^2(1 +\cos^2\theta +k'^2\cos^2\theta)$, and  $c=k'^4\cos^2\theta$. 
Derivatives of the plasma frame dispersion are required to evaluate the gradient in Equation \ref{eq:gradm} and can be written as
\begin{align}
\frac{\partial\mathcal{M}}{\partial k'}
= \frac{4A\,k'^3 + 2B\,k'}{\omega'^4(1-\omega'^2)},
\end{align}
\begin{align}
\frac{\partial\mathcal{M}}{\partial\omega'}
= \frac{2\left[a\,\omega'^6 + 2b\,\omega'^4 + (3c-b)\,\omega'^2 - 2c\right]}{\omega'^5\left(1-\omega'^2\right)^2},
\end{align}

\begin{align}
\frac{\partial\mathcal{M}}{\partial\theta}= -\frac{2\sin(\theta)\cos(\theta)[k'^4(1-\omega'^2) -\omega'^2k'^2]}{\omega'^4 (1-\omega'^2)},
\end{align}
In the HCI frame, which contains a background flow, the dispersion $\mathcal{M}_{HCI}$ is obtained by making the substitution $\omega_{HCI}=\omega+\mathbf{k}\cdot\mathbf{v}$ such that $\mathcal{M}_{HCI}(\omega_{HCI},\mathbf{k},\mathbf{v},\mathbf{r},t)=\mathcal{M}(\omega,\mathbf{k},\mathbf{r},t),\ \text{with}\ \omega=\omega_{HCI}-\mathbf{k}\cdot\mathbf{v}$. Following \cite{Hazeltine2004}, derivatives of  $\mathcal{M}_{HCI}$ are taken with respect to each of the independent variables $\mathbf{k}, \mathbf{r}, t, \omega$ of $\mathcal{M}_{HCI}$ are taken, holding each other variable constant, which results in the ray Equations \ref{eq:drdt_hci}, \ref{eq:dkdt_hci}, and \ref{eq:dkds_hci}. The relevant derivatives of $\mathcal{M}_{HCI}$ and $\mathcal{M}$
are related by 
\begin{align}
\frac{\partial\mathcal M_{HCI}}{\partial\omega}=\frac{\partial{\mathcal M}}{{\partial\omega}}\\
  \frac{\partial\mathcal M_{HCI}}{\partial \mathbf{r}}= \frac{\partial\mathcal M}{\partial \mathbf{r}}-\frac{\partial\mathcal M}{\partial \mathbf{\omega}}(\nabla\mathbf{v})\cdot\mathbf {k}\\
  \frac{\partial\mathcal M_{HCI}}{\partial \mathbf{k}}= \frac{\partial\mathcal M}{\partial \mathbf{k}}-\frac{\partial\mathcal M}{\partial \mathbf{\omega}}\mathbf{v}. \label{eq:gradm_hci}
  \end{align}

  \section{Comments on Error}
Error is introduced into this analysis in numerous places, though none of the results are severely affected by the error.

\paragraph{Dispersion Relation}Equation 1 is a cold-plasma dispersion relation which does not necessarily represent the true dispersion of the waves, which are relatively dispersive at ion-kinetic scales. A more accurate dispersion relation (e.g. hot-plasma with an empirical distribution function) \citep{ALPS}, would be ideal, though no closed form analytically differentiable term would likely be accessible. 
In the magnetosphere, hot-plasma tracing has been performed \citep{HorneThorne1993}. The hot-plasma solutions have similar form to the cold plasma dispersion at low-$\beta$ parallel propagating angles \citep{Klein2021,Mcmanus2024} and thus the error in the dispersion likely amounts to order unity corrections in derivatives of $\mathcal{M}$. The presence of a beam can affect the dispersion as well, though since the outward-going waves do not interact with the beam via standard cyclotron resonance, the outward waves are not strongly affected. Similarly the presence of $\alpha$-particles will modify the dispersion \citep{Mcmanus2024}, and these effects have been neglected as well. 
Note that the $\alpha$-population also contributes to the electron density $n_e$, which QTN measures. At nominal values of $4\%$ of the abundance \citep{Kasper2016}, their contribution is a $10\%$ error on the total density, though the effects on ray-tracing come through fractional densities, such that the effects from alpha abundance may not be particularly important in comparison to their effect on the dispersion.

\paragraph{Sampling Effects} Single spacecraft effects cannot fully resolve the full gradient fields that contribute to ray tracing and the measured gradients could presumably be much higher for some magnetic topologies, e.g. the gradient is mostly perpendicular to $B$ while the flow is mostly parallel to $B$. The gradient here is essentially a lower bound on the eikonal refraction, assuming that there are no other gradients (which seems nonphysical to say the least). The presence of enhanced unobserved gradients would increase the eikonal refraction, such that the turbulent gradients are even more dominant over the background field. If unobserved gradients are significantly larger, then it is possible that eikonal effects could become relatively more important with respect to the kinetic processes.

Conditioning on sampling angles is typically performed in studies of turbulence to understand anisotropy in gradients \citep{Horbury2012}. For the ray tracing, this could be performed to get an idea of the full 3D gradient field, which would also provide some more detailed information on the refraction of the waves. 

However, this conditional sampling also requires making assumptions about the homogeneity and stationarity of the system, which may not be valid since the gradients that correspond to large field inversions to enable a change in sampling direction are indeed the same gradients that one would hope to measure: i.e. the presence of large gradient that modifies the relative spacecraft sampling may result in a biased measurement of the gradients. In principle, simulation data could be used to better understand these effects.

There is also error introduced in the use of Taylor hypothesis. In this stream $V_{sc}/v_A \approx 5$, such that Alfv\'{e}nic fluctuations which are propagating at the background gradient scale at $\sim v_A$ introduce an error of 20\% in the gradient. This similarly affects the mapping of frequency to wavenumber with the $2\pi f_{sc} = \omega + kV_{sc}$ Doppler shift equation, with a similar fractional error. These 20\% estimates do not affect the order-of-magnitude scaling of the turbulent gradients versus the background Parker spiral, but care should be taken in precise wave-vector evolution via ray tracing. 

  \end{appendix}

 \end{document}